\documentclass[12pt, reqno]{amsart}
\usepackage{amsmath, amssymb, amsthm, epsfig}

\newtheorem{thm}{Theorem}[section]

\newtheorem{lemma}[thm]{Lemma}
\newtheorem{theorem}[thm]{Theorem}

\numberwithin{equation}{section}

\theoremstyle{definition}
\newtheorem{rem}[thm]{Remark}

\newcommand{\al}{\alpha}
\renewcommand{\b}{\beta}
\renewcommand{\c}{\gamma}

\newcommand{\e}{\varepsilon}

\renewcommand{\phi}{\varphi}

\renewcommand{\d}{\partial}

\renewcommand{\qed}{\rule{3mm}{3mm}}
\renewenvironment{proof}
{\vspace{1mm}\noindent\textbf{Proof.}} {\hspace*{\fill} $\qed$\vspace{1mm}}

\begin{document}
\title[Sharp bounds on $2m/r$]{Sharp bounds on $2m/r$ for static spherical objects}
\author{Paschalis Karageorgis}
\author{John G. Stalker}
\address{School of Mathematics, Trinity College, Dublin 2, Ireland.}
\email{pete@maths.tcd.ie}
\email{stalker@maths.tcd.ie}

\begin{abstract}
Sharp bounds are obtained, under a variety of assumptions on the eigenvalues of the Einstein tensor, for the ratio of the
Hawking mass to the areal radius in static, spherically symmetric space-times.
\end{abstract}
\maketitle

\section{Introduction}
All of the space-times considered in this paper are connected, four-dimensional and satisfy the following conditions.
\begin{itemize}
\item \emph{Spherical Symmetry.}\footnote{We follow Synge~\cite{Synge} in calling this assumption
``spherical symmetry'' for brevity.  This is a bit misleading, as we are assuming more than just spherical symmetry.  Our
assumption excludes, for example, Schwarzschild space, which lacks a time axis.} There is a time-like curve, called the
\emph{time axis}, with the property that at any point all normal directions are equivalent, \textit{i.e.}, that for any
two space-like normal unit vectors there is an isometry of the space-time which fixes the point and takes the first
vector to the second.  This defines an action of $SO(3)$ on the space-time whose orbits are called \emph{spheres}.

\item \emph{Staticity.}
There is a one-parameter group of isometries, called time translations, whose generating vector field is everywhere
time-like.

\item \emph{Regularity.}
The space-time, together with its metric, is of class $\mathcal{C}^3$, except possibly on 3-surfaces of discontinuity,
where the second derivatives of the metric are allowed to have jump discontinuities.
\end{itemize}

The somewhat odd looking regularity assumption is borrowed from Lichnerowicz~\cite{Lich}. It is meant to allow such
discontinuities as one expects to find at the interface between two different materials, but nothing worse.

The areal radius~$r$ is defined by the requirement that the area of a sphere be $4\pi r^2$. In terms of the radius~$r$
and metric tensor~$g$, we may then define the Hawking mass~$m$ by the relation
\begin{equation} \label{Hm}
g^{jk} \partial_j r \partial_ k r = 1 - \frac{2m}{r} \,.
\end{equation}
The purpose of this paper is to prove sharp bounds on the ratio $2m/r$ under various hypotheses on the eigenvalues of the
Einstein tensor. The particular hypotheses considered, their history and the resulting bounds are discussed in
Section~\ref{matter}.

Three general comments should be made at this stage.  First, the method employed is quite general and can be used to
obtain sharp bounds on $2m/r$ for any matter model, not just those described below. Second, obtaining sharp bounds is, in
each case, relatively easy.  Proving sharpness, while not conceptually difficult, requires considerably more effort.
Third, we carefully avoid the assumption, made tacitly by previous authors, that $2 m < r$. This point is discussed in
more detail in the next section.

Section~\ref{coordinates} is devoted to a discussion of coordinates and the components of the Einstein tensor in our
chosen coordinate system. Section~\ref{matter} introduces the various assumptions on this tensor which are needed for the
statement of our theorem. Our main result, Theorem~\ref{main}, appears in Section~\ref{omr}, while its proof is given in
Section~\ref{pro}.

\section{Geometry and Coordinates}\label{coordinates}
A space-time of the class considered above has coordinates $r$, $\theta$, $\phi$ and $t$, known as
\emph{curvature coordinates}, in which the metric takes the form
\begin{equation}\label{met}
g_{jk} \,dx^j dx^k = e^\al\,dr^2 + r^2(d\theta^2 + \sin^2\theta \,d\phi^2) - e^\c \,dt^2.
\end{equation}
Here, $\al$ and $\c$ are functions of $r$. As shown in Synge~\cite{Synge}, the Einstein tensor in curvature coordinates
is of the form
\begin{align}
G_r^r &= r^{-2} - r^{-2}e^{-\al} (1+r\c'), \label{eq1} \\
G_\theta^\theta = G_\phi^\phi &= e^{-\al} \left( -\frac{1}{2} \,\c'' -\frac{1}{4} \,\c'\c' - \frac{1}{2r} \,\c' +
\frac{1}{2r} \,\al' + \frac{1}{4} \,\al'\c' \right), \label{eq2}\\
G_t^t &= r^{-2} - r^{-2}e^{-\al} (1-r\al'). \label{eq3}
\end{align}
Here, primes denote derivatives with respect to $r$, while the off-diagonal entries are all zero.  The formulae become a
bit cleaner when one uses derivatives with respect to
\begin{equation*}
\b = 2\log r,
\end{equation*}
instead. Denoting such derivatives by dots, one obtains the equivalent system
\begin{align}
G_r^r &= r^{-2} - r^{-2}e^{-\al} (1+2\dot{\c}), \label{eq4} \\
G_\theta^\theta = G_\phi^\phi &= r^{-2} e^{-\al} (-2\ddot{\c} -\dot{\c}^2 +\dot{\al} + \dot{\al}\dot{\c}), \label{eq5}\\
G_t^t &= r^{-2} - r^{-2}e^{-\al} (1-2\dot{\al}). \label{eq6}
\end{align}
The corresponding Einstein tensor, given by Einstein's field equations, has diagonal entries
\begin{equation}\label{diag}
G_r^r = -8\pi p_R, \quad\quad G_\theta^\theta = G_\phi^\phi = -8\pi p_T, \quad\quad G_t^t = 8\pi \mu
\end{equation}
and all off-diagonal entries equal to zero.  Here, $p_R$ and $p_T$ are interpreted as the radial and tangential
pressures, respectively, while $\mu$ is interpreted as the energy density.

There are two annoying points about curvature coordinates.
\begin{itemize}
\item
The functions $\alpha$ and $\gamma$ may be of lower regularity than the metric, since $r$ itself is of lower regularity
than the metric.  This is discussed in more detail by Israel~\cite{Israel}.  For our purposes it suffices to note that
regularity is, in the presence of the other assumptions, equivalent to the statement that $\alpha$ and $\gamma$ are
$\mathcal{C}^3$ functions of~$r$, except possibly at certain points, where the radial pressure~$p_R$ is continuous and
the tangential pressure~$p_T$ and energy density $\mu$ are allowed to have jump discontinuities.  At $r = 0$, the correct
condition is that $\al'(0)= \c'(0) = 0$.

\item
The coordinates may fail to cover the whole space-time.  In fact, they cover the region from the time axis out to the
first marginally trapped sphere, \emph{i.e.}, the first sphere where $r = 2 m$.  If we were to assume, as most authors
do, that curvature coordinates cover the whole space-time, then we would, in effect, be making the very strong additional
assumption that $2 m / r < 1$ everywhere.  This we wish to avoid.  For the classes of space-times we consider it is, in
fact, the case that $ 2 m / r < 1$ everywhere, but this belongs to the conclusion of our theorem, not to its hypotheses.

The simplest example of a space-time that satisfies our spherical symmetry, staticity and regularity assumptions but has
a marginally trapped surface is de Sitter space, for which $e^\al = - e^{-\c} = 1 - r ^ 2 / R ^ 2$.  In this case, the
coordinates cover a region where $r < R$ but break down at the boundary.  Outside this region, there is another which is
isometric to the first, and it is easy to check that $2 m = r$ at $r = R$.  However, de Sitter space does not satisfy the
hypotheses of our theorem because it has negative pressures everywhere.
\end{itemize}

\section{Matter Models}\label{matter}
Various conditions on the three functions $p_R$, $p_T$ and $\mu$ are of interest:
\begin{itemize}
\item \textit{Non-negative Isotropic Pressure:}
For fluids, one expects $p_R = p_T \geq 0$.  The sharp bound
\begin{equation*}
2 m / r \leq 12\sqrt 2 - 16 \approx 0.9706
\end{equation*}
under this assumption, and no others, was derived by Bondi~\cite{Bondi}.  His method of proof is closely related to ours
but is not rigorous.

\item \textit{Buchdahl Assumption:}
For static stars with constant density, one has the bound
\begin{equation*}
2 m / r \leq 8 / 9
\end{equation*}
derived by Buchdahl \cite{Buch}.  More generally, this bound holds when $p_R= p_T\geq 0$, as long as $\mu\geq 0$ is
decreasing; see \cite{Buch}.  The isotropy assumption was relaxed in \cite{GuMu}, where the case $p_R\geq p_T\geq 0$ was
treated, still the monotonicity assumption remains crucial.

\item \textit{Dominant Energy Condition:}
For almost any reasonable matter model, one expects $|p_R| \leq \mu$ and $|p_T| \leq \mu$.  In the special case that
$p_R,p_T\geq 0$, the bound
\begin{equation*}
2 m / r \leq 48 / 49 \approx 0.9796
\end{equation*}
is provided by \cite{Hak}. Our bound for this special case is roughly $0.963$, which we show to be sharp.

\item \textit{Vlasov-Einstein:}
For Vlasov-Einstein matter, the stress energy tensor is an integral of those of individual particles, each of which has
rank one and satisfies the dominant energy condition.  This implies that $p_R\geq 0$, $p_T \geq 0$ and $p_R + 2p_T \leq
\mu$. Under these assumptions, Andr\'easson \cite{Hak} has recently shown that the sharp\footnote{A somewhat unfortunate
feature of our argument is that the sharpness of the estimate $\frac{2m}{r} \leq \frac{8}{9}$ is proved only within the class
of space-times satisfying the pressure conditions above, without considering whether such space-times arise from solutions of
the full Vlasov-Einstein system. Andr\'easson's argument, on the other hand, does provide solutions to the full system.} bound
is
\begin{equation*}
2 m / r \leq 8 / 9.
\end{equation*}
Our method provides a new, and considerably shorter, proof of this result.

\item \textit{Zero Radial Pressure:}
The case $p_T \geq p_R = 0$ was studied by Florides~\cite{Flo} who obtained the sharp bound
\begin{equation*}
2 m / r \leq 2 / 3.
\end{equation*}
This can also be proved using our method, but the resulting proof is neither shorter nor clearer than the original, so we
do not consider this case further.
\end{itemize}

\section{Our main result}\label{omr}
\begin{theorem}\label{main}
Consider a space-time satisfying the regularity, staticity and spherical symmetry conditions described in the
introduction. Suppose that the corresponding Hawking mass \eqref{Hm} is finite and that the pressures $p_R,p_T$ and
energy density $\mu$ are all non-negative.

\begin{itemize}
\item[(1)] \textbf{Vlasov-Einstein case}. Assuming that $p_R +2p_T\leq \mu$, one has
\begin{equation}\label{es1}
\left( 4 - \frac{6m}{r} + 8\pi r^2 p_R\right)^2 \leq 16\left( 1- \frac{2m}{r} \right),
\quad\quad \frac{2m}{r} \leq \frac{8}{9} \,.
\end{equation}

\item[(2)] \textbf{Isotropic case}. Assuming that $p_R= p_T$, one has
\begin{equation}\label{es2}
\left( 2 - \frac{2m}{r} + 8\pi r^2 p_R\right)^2 \leq 36\left( 1- \frac{2m}{r} \right),
\quad\quad \frac{2m}{r} \leq 12\sqrt 2-16.
\end{equation}

\item[(3)] \textbf{Isotropic case with dominant energy}. Assuming that $p_R = p_T\leq \mu$, one has
\begin{equation}\label{es3}
\left( 2 - \frac{2m}{r} + 8\pi r^2 p_R\right)^2 \leq 9.551 \left( 1- \frac{2m}{r} \right),
\quad\quad \frac{2m}{r} \leq 0.865.
\end{equation}

\item[(4)] \textbf{Dominant energy in tangential direction}. Assuming that $p_T \leq \mu$, one has
\begin{equation}\label{es4}
\left( 6 - \frac{10m}{r} + 8\pi r^2 p_R\right)^2 \leq 40\left( 1- \frac{2m}{r} \right),
\quad\quad \frac{2m}{r} \leq \frac{2\sqrt 2+2}{5} \,.
\end{equation}

\item[(5)] \textbf{Dominant energy case}. Assuming that $p_R, p_T \leq \mu$, one has
\begin{equation}\label{es5}
\left( 6 - \frac{10m}{r} + 8\pi r^2 p_R\right)^2 \leq 37.924\left( 1- \frac{2m}{r} \right),
\quad\quad \frac{2m}{r} \leq 0.963.
\end{equation}
\end{itemize}
Moreover, these ten estimates are all sharp for the class of space-times considered in each case.  As for the numerical
values that appear in \eqref{es3} and \eqref{es5}, these can be described in terms of a system of ODEs which arises in
the course of the proof; see \eqref{ode}.  The values given here are accurate up to three decimal places.
\end{theorem}

\begin{rem}
There is no assumption on the behavior of the space-time as $r$ tends to infinity.  In fact, we do not even assume that
$r$ is unbounded. This point is crucial.  It allows us to apply the theorem to the interior of a finite sphere and, in
particular, to the interior of the first marginally trapped surface, if there is such a surface.

More precisely, suppose we can prove the theorem in the region where the curvature coordinates are defined, namely in the
region where $2 m < r$. For each matter model, we may then deduce that $2m/r \leq c < 1$ for some constant $c$ which
depends on the matter model considered. Since $2 m/r$ is continuous and our space-time is connected, this actually
implies that $2m/r \leq c$ throughout the space-time. In other words, the marginally trapped surface that we allowed is
not, in fact present, and the curvature coordinates, which might \emph{a priori} have covered only part of the
space-time, cover the whole space-time. We therefore obtain the full theorem from the special case where the whole
space-time is covered by curvature coordinates.  In particular, we may, and do, use curvature coordinates throughout the
proof without further comment.
\end{rem}

\begin{rem}
Our proof for the isotropic case $p_R = p_T$ applies verbatim in the more general case $p_R \geq p_T$, while the
estimates \eqref{es2} are sharp for that case as well.
\end{rem}

\begin{rem}
The \textit{assumptions} of Theorem \ref{main} can be slightly improved in the sense that we do not use our hypothesis
$p_T\geq 0$ to establish the given estimates.  This hypothesis is merely included to improve the \textit{conclusions} of
Theorem \ref{main}, as sharpness is now shown over a smaller class of space-times. In fact, the examples we construct in
order to prove sharpness belong to the even smaller class of space-times which are vacuum outside a sphere.
\end{rem}

The proof of Theorem \ref{main} is based on the following elementary fact, which is essentially due to Bondi
\cite{Bondi}.

\begin{lemma}\label{cur}
Let the assumptions of Theorem \ref{main} hold. Then the variables
\begin{equation} \label{c}
x \equiv 1-e^{-\al} = \frac{2m}{r} \,,\quad\quad y \equiv -r^2 G_r^r = 8\pi r^2 p_R
\end{equation}
give rise to a parametric curve which lies in $[0,1)\times [0,\infty)$ and satisfies the equations
\begin{align}
8\pi r^2 p_R &= y, \label{cur1} \\
8\pi r^2 p_T &= \frac{x+y}{2(1-x)} \:\dot{x} + \dot{y} + \frac{(x+y)^2}{4(1-x)} \,, \label{cur2} \\
8\pi r^2 \mu &= 2\dot{x} + x, \label{cur3}
\end{align}
where the dots denote derivatives with respect to $\b= 2\log r$.
\end{lemma}

\begin{proof}
First of all, we combine equations \eqref{eq3} and \eqref{diag} to write
\begin{equation*}
8\pi r^2 \mu = 1 - \d_r (re^{-\al}).
\end{equation*}
Integrating over $[0,r]$ and using the definition of the Hawking mass \eqref{Hm}, we then get
\begin{equation} \label{nd1}
\frac{2m}{r} = 1 -e^{-\al} = x.
\end{equation}
This implies $x\geq 0$ because $m\geq 0$ whenever $\mu\geq 0$.  Next, we use \eqref{diag} to get
\begin{equation*}
y \equiv - r^2 G_r^r = 8\pi r^2 p_R \geq 0.
\end{equation*}
To establish our assertion \eqref{cur3}, we combine \eqref{diag}, \eqref{eq6} and \eqref{nd1} to find that
\begin{equation*}
8\pi r^2 \mu = r^2 G_t^t = 1 - e^{-\al} (1-2\dot{\al}) = 1 - (1-x) \left( 1- \frac{2\dot{x}}{1-x} \right) = x+2\dot{x}.
\end{equation*}
To establish our remaining assertion \eqref{cur2}, we first use \eqref{eq4} and \eqref{nd1} to get
\begin{equation*}
y\equiv -r^2 G_r^r = -1 + (1-x)(1+2\dot{\c}), \quad\quad \dot{\al} = \frac{\dot{x}}{1-x} \,.
\end{equation*}
Solving the leftmost equation for $\dot{\c}$ and differentiating, we conclude that
\begin{equation*}
2\dot{\c} = \frac{x+y}{1-x} \,,\quad\quad 2\ddot{\c} = \frac{(1+y) \dot{x} + (1-x) \dot{y}}{(1-x)^2} \,.
\end{equation*}
On the other hand, equations \eqref{diag}, \eqref{eq5} and \eqref{nd1} combine to give
\begin{equation*}
8\pi r^2 p_T = (1-x)(2\ddot{\c} +\dot{\c}^2 -\dot{\al} - \dot{\al}\dot{\c}).
\end{equation*}
Using these facts and a simple computation, one may thus easily deduce \eqref{cur2}.
\end{proof}

\section{Proof of Theorem \ref{main}} \label{pro}
To prove the desired estimates, we study the curve \eqref{c} provided by Lemma \ref{cur}.  In each case, we are seeking
an upper bound for $x= 2m/r$ and also an upper bound for
\begin{equation}\label{wn}
w_n(x,y) = \frac{(n(1-x)+1+y)^2}{1-x}\,,
\end{equation}
where the exact value of $n$ varies from case to case. Differentiating \eqref{wn}, we get
\begin{equation}\label{wd}
\dot{w}_n = \frac{n(1-x)+1+y}{(1-x)^2} \cdot \Bigl[ (1+y-n(1-x)) \dot{x} + 2(1-x) \dot{y} \Bigr]
\end{equation}
throughout the curve \eqref{c}, where dots denote derivatives with respect to $\b= 2\log r$.  In the special case that
$n=1$, this formula reads
\begin{equation}\label{w1d}
\dot{w}_1 = \frac{2-x+y}{(1-x)^2} \cdot \Bigl[ (x+y) \dot{x} + 2(1-x) \dot{y} \Bigr]
\end{equation}
and it is closely related to the tangential pressure $p_T$; see \eqref{cur2}.  Let us also recall that
\begin{equation*}
0\leq x < 1, \quad\quad y\geq 0
\end{equation*}
throughout the curve \eqref{c}, a fact we shall frequently need to use in what follows.

\subsection{Vlasov-Einstein case}
In this case, we are assuming that $p_R + 2p_T\leq \mu$.  According to Lemma \ref{cur}, the corresponding curve \eqref{c}
must thus satisfy
\begin{align}\label{1-as}
(3x+y-2) \dot{x} + 2(1-x) \dot{y} \leq -\frac{z_3(x,y)}{2}\,,\quad\quad z_3 = 3x^2 - 2x + y^2 + 2y.
\end{align}
Combining the last equation with our computation \eqref{wd}, we now find
\begin{align*}
\dot{w}_3 &= \frac{4-3x+y}{(1-x)^2} \cdot \Bigl[ (3x+y-2) \dot{x} + 2(1-x) \dot{y} \Bigr] \\
&\leq -\frac{4-3x+y}{2(1-x)^2} \cdot z_3(x,y).
\end{align*}
In particular, $w_3$ is decreasing whenever $z_3> 0$, so it must be the case that
\begin{equation*}
w_3 \leq \max_{\substack{0\leq x\leq 1\\ z_3\leq 0\leq y}} w_3(x,y) = w_3(0,0) = 16
\end{equation*}
throughout the curve.  This proves the first inequality in \eqref{es1}, which also implies the second inequality because
the maximum value of $x$ over the region $0\leq x\leq 1$, $y\geq 0$, $w_3\leq 16$ is attained at $(8/9,0)$, namely at the
point at which the curve $w_3=16$ intersects the $x$-axis. We refer the reader to Fig. 1 for a sketch of the curves
$z_3=0$ and $w_3=16$.

To show that the estimates in \eqref{es1} are sharp, we need to construct a space-time such that the corresponding curve
of Lemma \ref{cur} intersects a small neighbourhood of $(8/9,0)$.  Let us now temporarily assume that we have a
parametric curve
\begin{equation*}
x= x(\tau), \quad\quad y= y(\tau); \quad\quad \tau\in (0,\infty)
\end{equation*}
which passes near the point $(8/9,0)$ and also satisfies the following properties:
\begin{itemize}
\item[(A1)] $\frac{1}{z_3} \cdot \frac{dw_3}{d\tau}$ is both negative and integrable;
\item[(A2)] $0\leq x(\tau) < 1$ and $y(\tau)\geq 0$ for each $\tau > 0$;
\item[(A3)] $y(\tau)=0$ for all large enough $\tau$ and $x(\tau)\to 0$ as $\tau\to\infty$;
\item[(A4)] the curve is $\mathcal{C}^1$ except for finitely many points.
\end{itemize}
Given such a curve, we can easily construct a space-time as follows. First, we define
\begin{equation}\label{def1}
\kappa(\tau) = -\frac{dw_3/d\tau}{z_3(x,y)} \cdot \frac{2(1-x)^2}{4-3x+y}
\end{equation}
and we note that $\kappa$ is both positive and integrable by (A1)-(A2). Next, we define
\begin{equation}\label{def2}
\b = \int \kappa\:d\tau, \quad\quad r= \exp(\b/2)
\end{equation}
and finally, we define the metric coefficients in \eqref{met} by
\begin{equation}\label{def3}
\al(r) = -\log(1-x), \quad\quad \c(r)= \int \frac{x+y}{2(1-x)} \cdot \kappa\,d\tau.
\end{equation}
Letting dots denote derivatives with respect to $\b= 2\log r$, as usual, we then get
\begin{equation*}
\dot{w}_3 = \frac{1}{\kappa} \cdot \frac{dw_3}{d\tau} = -\frac{4-3x+y}{2(1-x)^2} \cdot z_3(x,y)
\end{equation*}
using our definitions \eqref{def2} and \eqref{def1}.  In view of our computation \eqref{wd}, this gives
\begin{equation}\label{1-def}
(3x+y-2) \dot{x} + 2(1-x) \dot{y} = -\frac{z_3(x,y)}{2} \,,
\end{equation}
which is equivalent to the equation $p_R + 2p_T = \mu$ because of Lemma \ref{cur}.

To finish the proof for this case, it thus remains to construct the curve whose existence we assumed in the previous
paragraph. We have to ensure that the curve satisfies (A1)-(A4), that it passes arbitrarily close to $(8/9,0)$ and that
the corresponding quantities $p_R,p_T, \mu$ provided by Lemma \ref{cur} are non-negative.  Let us then fix some small
$\e>0$ and consider the curve
\begin{equation}\label{1-A}
w_{3-3\e}(x,y) = \Bigl[ \e \sqrt{1+3x} + 4(1-\e) \Bigr]^2.
\end{equation}
When $\e=0$, this reduces to the curve $w_3=16$ which passes through the origin and $(8/9,0)$.  When $\e=1$, it reduces
to the curve $z_3=0$ which passes through the origin and $(2/3,0)$.  In the more general case $0<\e<1$, it describes a
curve that lies between these two curves.  We start out at the origin and follow this curve until we hit the $x$-axis,
and then we return to the origin along the $x$-axis.  Let us henceforth denote by $C_1$ the curve obtained in this
manner; we refer the reader to Fig. 1 for a typical sketch of this curve.

\begin{figure}[t]
\centerline{\psfig{figure=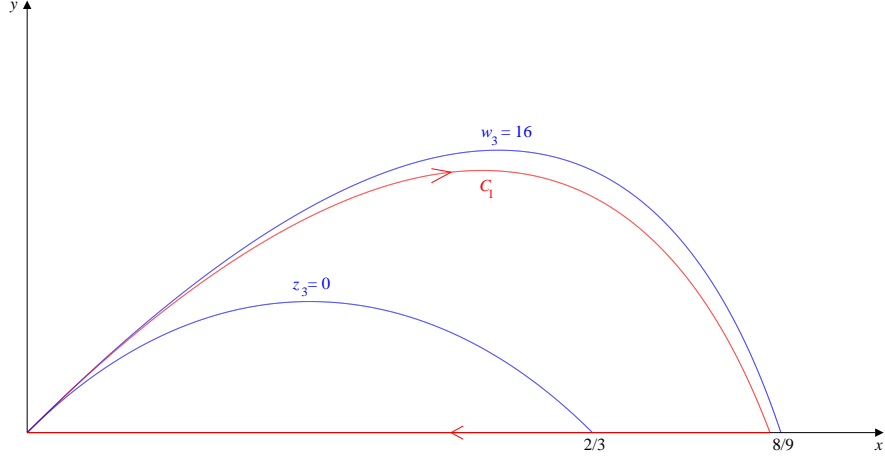,height=6cm}\psfull}
\caption{The curve $C_1$ for the Vlasov-Einstein case.}
\end{figure}

The fact that $C_1$ satisfies (A2)-(A4) is trivial.  To check that it satisfies (A1) along the part defined by
\eqref{1-A}, we recall that this part lies between the curves $w_3=16$ and $z_3=0$.  Thus, it is easy to see that $z_3>0$
along this part, and we need only check that
\begin{equation} \label{ch1}
\frac{d w_3}{d \tau} < 0
\end{equation}
as one follows the curve \eqref{1-A} in the positive $x$-direction. Differentiation of \eqref{1-A} gives
\begin{align*}
\frac{dw_{3-3\e}}{d\tau} &= \frac{3\e \sqrt{w_{3-3\e}}}{\sqrt{1+3x}} \cdot \frac{dx}{d\tau} \\
&= \frac{3\e}{\sqrt{1+3x}} \cdot \frac{3(1-\e)(1-x)+1+y}{\sqrt{1-x}} \cdot \frac{dx}{d\tau}
\end{align*}
along the curve \eqref{1-A}, and we may compare this equation with \eqref{wd} to find that
\begin{equation} \label{1-Ad}
2(1-x) \cdot \frac{dy}{d\tau} = \left[ \frac{3\e(1-x)^{3/2}}{\sqrt{1+3x}} + 3(1-\e)(1-x) -1-y \right] \cdot
\frac{dx}{d\tau}
\end{equation}
along the curve \eqref{1-A}.  Employing our computation \eqref{wd} once again, we deduce that
\begin{equation*}
\frac{dw_3}{d\tau} = \frac{3\e(4-3x+y)}{1-x} \cdot
\frac{\sqrt{1-x}-\sqrt{1+3x}}{\sqrt{1+3x}} \cdot \frac{dx}{d\tau} \,.
\end{equation*}
Since $\frac{dx}{d\tau}>0$ here, the desired \eqref{ch1} follows.  To show that (A1) also holds for the remaining part of
the curve $C_1$, we need only note that
\begin{equation*}
\frac{1}{z_3} \cdot \frac{d w_3}{d \tau} = \frac{1}{x} \cdot \frac{4-3x}{(1-x)^2} \cdot \frac{dx}{d\tau} < 0
\end{equation*}
along the line $y=0$ because this line is traversed in the direction of decreasing $x$.

Finally, we check that $p_R,p_T,\mu\geq 0$ throughout the curve $C_1$. The fact that $p_R\geq 0$ follows by (A2) because
$8\pi r^2 p_R = y$ by definition.  Since \eqref{1-def} ensures that $\mu = p_R + 2p_T$, we need only check that $p_T\geq
0$ as well. Let us now write
\begin{align}
8\pi r^2 p_T &= \frac{x+y}{2(1-x)} \:\dot{x} + \dot{y} + \frac{(x+y)^2}{4(1-x)} \notag \\
&= \frac{1-x}{2(2-x+y)} \:\dot{w}_1 + \frac{(x+y)^2}{4(1-x)} \label{pT}
\end{align}
using equations \eqref{cur2} and \eqref{w1d}. Along the part of $C_1$ defined by \eqref{1-A}, we have
\begin{equation*}
\dot{w}_1 = \frac{2-x+y}{1-x} \cdot \left[ \frac{3\e\sqrt{1-x}}{\sqrt{1+3x}} + 2-3\e \right] \cdot \dot{x}
\end{equation*}
by \eqref{w1d} and \eqref{1-Ad}.  In view of our definition \eqref{def2}, we thus have
\begin{equation*}
\dot{w}_1 = \frac{2-x+y}{1-x} \cdot \left[ \frac{3\e\sqrt{1-x}}{\sqrt{1+3x}} + 2-3\e \right]
\cdot \frac{dx/d\tau}{\kappa} \,.
\end{equation*}
Since $\e>0$ is small and $\kappa$ is positive by above, this implies $\dot{w}_1>0$, hence $p_T>0$ by \eqref{pT}.  For
the remaining part of $C_1$ along the $x$-axis, Lemma \ref{cur} and \eqref{1-def} give
\begin{equation*}
p_R= 0, \quad\quad p_T = \frac{x\mu}{4(1-x)} \,,\quad\quad \mu = 2p_T,
\end{equation*}
so it easily follows that $p_R= p_T = \mu = 0$ throughout this part of the curve.

\subsection{Isotropic case} In this case, our assumption that $p_R =p_T$ is equivalent to
\begin{align}\label{2-as}
(x+y) \dot{x} + 2(1-x) \dot{y} = - \frac{z_1(x,y)}{2} \,,\quad\quad  z_1 = (x+y)^2 -4y(1-x).
\end{align}
Proceeding as before, we use our computation \eqref{w1d} to find that
\begin{align} \label{2-m}
\dot{w}_1 = \frac{2-x+y}{(1-x)^2} \cdot \Bigl[ (x+y) \dot{x} + 2(1-x) \dot{y} \Bigr]= -\frac{2-x+y}{2(1-x)^2}
\cdot z_1(x,y).
\end{align}
Once again, $w_1$ is decreasing as soon as $z_1> 0$, so it must be the case that
\begin{equation*}
w_1 \leq \max_{\substack{0\leq x\leq 1\\ z_1\leq 0\leq y}} w_1(x,y) = w_1(0,4) = 36
\end{equation*}
throughout the curve.  This proves the first inequality in \eqref{es2}, while the second inequality follows because the
maximum value of $x$ over the region $0\leq x\leq 1$, $y\geq 0$, $w_1\leq 36$ is attained at $(12\sqrt 2-16,0)$.

To show that the estimates in \eqref{es2} are sharp, we argue as in the previous case.  Suppose we have a curve which
passes near the point $(12\sqrt 2-16,0)$ and satisfies
\begin{itemize}
\item[(B1)] $\frac{1}{z_1} \cdot \frac{dw_1}{d\tau}$ is both negative and integrable
\end{itemize}
as well as (A2)-(A4).  Then we can follow our previous approach with
\begin{equation}\label{def4}
\kappa(\tau) = -\frac{dw_1/d\tau}{z_1(x,y)} \cdot \frac{2(1-x)^2}{2-x+y} >0
\end{equation}
instead of \eqref{def1}. Our definitions \eqref{def2}-\eqref{def3} are still applicable, however they now imply
\begin{equation} \label{def5}
\dot{w}_1 = \frac{1}{\kappa} \cdot \frac{dw_1}{d\tau} = -\frac{2-x+y}{2(1-x)^2} \cdot z_1(x,y).
\end{equation}
In view of our computation \eqref{w1d}, they thus imply
\begin{equation}\label{2-def}
(x+y) \dot{x} + 2(1-x) \dot{y} = -\frac{z_1(x,y)}{2} \,,
\end{equation}
which is equivalent to the equation $p_R = p_T$ because of Lemma \ref{cur}.

To finish the proof for this case, it thus remains to construct the curve whose existence we assumed in the previous
paragraph. Fix some small $\e>0$ and set
\begin{equation}\label{2-xy}
x_\e= \e, \quad\quad y_\e=2-3\e + 2\sqrt{(1-\e)(1-2\e)}
\end{equation}
for convenience.  Then $(x_\e,y_\e)$ is a point on the curve $z_1=0$ which is close to $(0,4)$. To define the first part
of the desired curve, we use the equation
\begin{align}\label{2-A}
\sqrt{w_1(x,y)} &= \sqrt{w_1(0,0)} + 2A_\e x_\e x - A_\e x^2,
\end{align}
where $A_\e$ is determined by requiring that the curve passes through $(x_\e, y_\e)$, namely
\begin{equation}\label{2-AB}
A_\e = \frac{ \sqrt{w_1(x_\e,y_\e)} - 2}{x_\e^2} \,.
\end{equation}
We start out at the origin and we follow the curve \eqref{2-A} until we reach the point $(x_\e,y_\e)$, then we follow the
curve
\begin{equation} \label{2-B}
\sqrt{w_1(x,y)} = \sqrt{w_1(x_\e,y_\e)} - \frac{\e (x-x_\e)^2}{\sqrt{1-x}} \,,\quad\quad x\geq x_\e
\end{equation}
until we hit the $x$-axis, and finally we return to the origin along the $x$-axis.  Let $C_2$ denote the curve obtained
in this manner; a typical sketch of this curve appears in Fig. 2.

\begin{figure}[t] \centerline{\psfig{figure=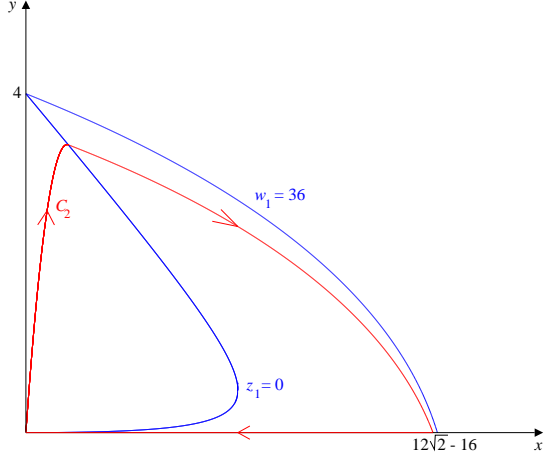,height=6cm}\psfull}
\caption{The curve $C_2$ for the isotropic case.}
\end{figure}

The fact that $C_2$ satisfies (A2)-(A4) is trivial; we now check that it satisfies (B1).  When it comes to the part of
$C_2$ defined by \eqref{2-A}, we have $z_1\leq 0$, $x\leq x_\e$ and
\begin{equation}\label{2-Ad}
\frac{dw_1/d\tau}{2\sqrt{w_1}} = 2A_\e (x_\e - x) \cdot \frac{dx}{d\tau} \,.
\end{equation}
Since $x$ is increasing along this part of $C_2$, it thus suffices to check that $A_\e$ is positive.  In view of
\eqref{2-AB}, this is certainly the case for all small enough $\e>0$ because
\begin{align*}
\lim_{\e \to 0} \: \e^2 A_\e &= \sqrt{w_1(0,4)} - 2 = 4.
\end{align*}
When it comes to the part of $C_2$ defined by \eqref{2-B}, we have $z_1\geq 0$, $x\geq x_\e$ and
\begin{equation*}
\frac{dw_1/d\tau}{2\sqrt{w_1}} = - \e (x-x_\e) \cdot \frac{4(1-x)+x-x_\e}{2(1-x)^{3/2}} \cdot
\frac{dx}{d\tau} \leq 0,
\end{equation*}
as needed.  When it comes to the remaining part of $C_2$ along the $x$-axis, we have
\begin{equation*}
\frac{dw_1/d\tau}{z_1} = \frac{2-x}{x(1-x)^2} \cdot \frac{dx}{d\tau} <0,
\end{equation*}
and this shows that the desired property (B1) holds throughout the curve $C_2$.

Finally, we check that $p_R,p_T,\mu\geq 0$ throughout the curve $C_2$. The fact that $p_R\geq 0$ follows trivially as
before, hence $p_T=p_R\geq 0$ by \eqref{2-def} and we need only worry about $\mu$.  Since
\begin{equation*}
8\pi r^2 \mu = 2\dot{x} + x
\end{equation*}
by \eqref{cur3}, we have $\mu\geq 0$ as long as $x$ is increasing along the curve, so we need only check the part of
$C_2$ along the $x$-axis.  As in the previous case, however, Lemma \ref{cur} and \eqref{2-def} combine to give $p_R= p_T
= \mu = 0$ throughout this part, so the proof for this case is complete.

\subsection{Isotropic case with dominant energy} Our assumption that $p_R = p_T\leq \mu$ gives
\begin{equation} \label{3-as}
(x+y) \dot{x} + 2(1-x) \dot{y} = - \frac{z_1(x,y)}{2} \,,\quad\quad 2\dot{x} \geq y-x
\end{equation}
with $z_1$ as in \eqref{2-as}.  Due to the isotropy condition, \eqref{2-m} remains valid, so $w_1$ is increasing if and
only if $z_1 \leq 0$.  Since the curve of Lemma \ref{cur} starts out at the origin, where $z_1=0$, it may only attain the
largest possible value of $w_1$ at a point along the curve $z_1=0$. It is easy to check that higher values of $w_1$ occur
at higher points on this curve.  To attain the largest possible value of $w_1$, the curve of Lemma \ref{cur} must thus
ascend as fast as possible within the region $z_1\leq 0$.  Since it starts out at the origin, it must satisfy
\begin{equation*}
(x+y) \dot{x} + 2(1-x) \dot{y} = - \frac{z_1(x,y)}{2} \,,\quad\quad 2\dot{x} = y-x
\end{equation*}
until it exits the region $z_1\leq 0$.  This gives rise to the system of ODEs
\begin{equation}\label{ode}
2\dot{x} = y-x, \quad\quad 2\dot{y} = \frac{y(2-3x-y)}{1-x}
\end{equation}
which has a saddle point at the origin.  The solution of interest is the one corresponding to the unstable manifold
associated with the origin.  Using numerical integration, we find that it intersects the curve $z_1=0$ at
the point $(x_1,y_1) = (0.4927,0.6939)$; see Fig. 3.  This makes 
\begin{equation*}
w_1(x_1,y_1) \approx 9.551
\end{equation*}
the largest possible value of $w_1$, and then we can use the fact that $w_1\leq 9.551$ to deduce that the largest
possible value of $x$ is attained at $(0.865,0)$.

To show that our results for this case are sharp, we need to find a curve which passes near the point $(0.865,0)$ and
satisfies (B1) as well as (A2)-(A4).  Given such a curve, one can use our approach in the previous case to obtain a
space-time for which $p_R = p_T$. We start out at the origin and we follow the solution to the ODE
\begin{equation}\label{ode2}
\frac{dy}{dx} =\frac{y(2-3x-y)}{(1-x)(y-x)}
\end{equation}
corresponding to the associated unstable manifold; we do so until we reach the point $(x_1,y_1)$ that lies on the curve
$z_1=0$, then we follow the curve
\begin{equation} \label{3-B}
\sqrt{w_1(x,y)} = \sqrt{w_1(x_1,y_1)} - \frac{\e (x-x_1)^2}{\sqrt{1-x}} \,,\quad\quad x\geq x_1
\end{equation}
until we hit the $x$-axis, and finally we return to the origin along the $x$-axis. We refer the reader to Fig. 3 for a
typical sketch of the curve $C_3$ obtained in this manner.

\begin{figure}[t]
\centerline{\psfig{figure=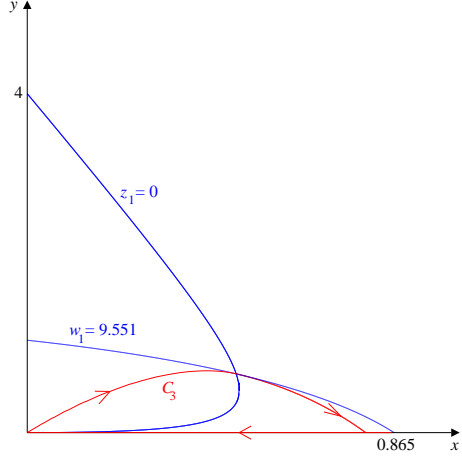,height=6cm}\psfull}
\caption{The curve $C_3$ for the isotropic case with dominant energy.}
\end{figure}

The only nontrivial properties we need to verify are (B1) and the fact that $p_R\leq \mu$.  When it comes to the part of
$C_3$ defined by \eqref{ode2}, we have $p_R= p_T= \mu$ and also
\begin{equation} \label{3def}
\frac{1}{z_1} \cdot \frac{d w_1}{d \tau} = -\frac{2-x+y}{(1-x)^2(y-x)} \cdot \frac{dx}{d\tau} \leq 0,
\end{equation}
so the desired properties are easily seen to hold.  The same is true for the part of $C_3$ along the $x$-axis because
$p_R= p_T = \mu = 0$ and since
\begin{equation*}
\frac{1}{z_1} \cdot \frac{d w_1}{d \tau} = \frac{1}{x} \cdot \frac{2-x}{(1-x)^2} \cdot \frac{dx}{d\tau} < 0
\end{equation*}
along this part.  For the remaining part defined by \eqref{3-B}, we have
\begin{equation}\label{3-dw1}
\frac{dw_1/d\tau}{2\sqrt{w_1}} = - \e (x-x_1) \cdot
\frac{4(1-x)+x-x_1}{2(1-x)^{3/2}} \cdot \frac{dx}{d\tau} \leq 0,
\end{equation}
which implies property (B1) because $z_1\geq 0$ for this part.  Writing \eqref{def5} in the form
\begin{equation*}
\dot{w}_1 = -\frac{2-x+y}{2(1-x)^2} \cdot z_1(x,y) = -\frac{\sqrt{w_1(x,y)}}{2(1-x)^{3/2}} \cdot z_1(x,y),
\end{equation*}
we now combine the last two equations to deduce that
\begin{equation*}
2\dot{x} = \frac{z_1(x,y)}{\e(x-x_1)(4-3x-x_1)}
\end{equation*}
throughout the curve \eqref{3-B}.  According to Lemma \ref{cur}, the condition $p_R\leq \mu$ we need to verify is
equivalent to the condition $2\dot x \geq y-x$, so we need to check that
\begin{equation} \label{ch2}
\frac{z_1(x,y)}{x-x_1} \geq \e(4-3x-x_1) (y-x)
\end{equation}
throughout the curve \eqref{3-B}.  Write equation \eqref{3-B} in the equivalent form
\begin{equation*}
y = f(x) \equiv \sqrt{w_1(x_1,y_1)} \sqrt{1-x} + x- 2 - \e (x-x_1)^2.
\end{equation*}
Then $z_1(x_1,f(x_1))= z_1(x_1,y_1)= 0$ by construction, so one easily finds
\begin{align*}
\lim_{x\to x_1} \frac{z_1(x,f(x))}{x-x_1} &= 8(x_1+y_1) - 4 - (3x_1+y_1-2)
\cdot \frac{\sqrt{w_1(x_1,y_1)}}{\sqrt{1-x_1}} \approx 4.746
\end{align*}
using \eqref{2-as}.  Thus, the left hand side of \eqref{ch2} is bounded away from zero near $x=x_1$.  Since the same is
true away from $x=x_1$, where $z_1$ itself is bounded away from zero, we can always find a small enough $\e>0$ so that
\eqref{ch2} holds throughout the curve \eqref{3-B}.

\subsection{Dominant energy in tangential direction} Our assumption that $p_T\leq \mu$ gives
\begin{align}\label{4-as}
(5x+y-4) \dot{x} + 2(1-x) \dot{y} \leq -\frac{z_5(x,y)}{2}\,,\quad\quad z_5 = (x+y)^2 - 4x(1-x).
\end{align}
Proceeding as before, we use our computation \eqref{wd} to find that
\begin{align} \label{4-m}
\dot{w}_5 = \frac{6-5x+y}{(1-x)^2} \cdot \Bigl[ (5x+y-4) \dot{x} + 2(1-x) \dot{y} \Bigr] \leq
-\frac{6-5x+y}{2(1-x)^2} \cdot z_5(x,y).
\end{align}
Once again, $w_5$ is decreasing as soon as $z_5> 0$, so it must be the case that
\begin{equation*}
w_5 \leq \max_{\substack{0\leq x\leq 1\\ z_5\leq 0\leq y}} w_5(x,y) = w_5(1/10,1/2) = 40
\end{equation*}
throughout the curve. This proves the first inequality in \eqref{es4}, and the second inequality follows as before.

To show that the estimates in \eqref{es4} are sharp, we need to find a curve which satisfies
\begin{itemize}
\item[(C1)] $\frac{1}{z_5} \cdot \frac{dw_5}{d\tau}$ is both negative and integrable
\end{itemize}
as well as (A2)-(A4).  Given such a curve, one can use our previous approach to obtain a space-time for which $p_T =
\mu$.  To define the first part of the curve, we use the equation
\begin{align} \label{4-A}
\sqrt{w_5(x,y)} = \sqrt{w_5(0,0)} + \frac{Ax}{5} - Ax^2,
\end{align}
where $A$ is chosen so that the curve passes through $(1/10,1/2)$, namely
\begin{equation*}
A= 100(\sqrt{40} -6) >0.
\end{equation*}
We start out at the origin and we follow the curve \eqref{4-A} until we reach the point $(1/10,1/2)$, then we follow the
curve
\begin{equation*}
\sqrt{w_5(x,y)} = \sqrt{w_5(1/10,1/2)} - \frac{\e (x-1/10)^2}{\sqrt{1-x}} \,,\quad\quad x\geq 1/10
\end{equation*}
until we hit the $x$-axis, and finally we return to the origin along the $x$-axis.  Since this curve is almost identical
with the one for the isotropic case, our previous approach applies with minor changes; we shall not bother to include the
details here.

\subsection{Dominant energy case} In this case, our assumption that $p_R,p_T\leq \mu$ gives
\begin{equation} \label{5-as}
(5x+y-4) \dot{x} + 2(1-x) \dot{y} \leq -\frac{z_5(x,y)}{2}\,,\quad\quad 2\dot{x} \geq y-x
\end{equation}
with $z_5$ as in \eqref{4-as}.  Since \eqref{4-m} remains valid, $w_5$ is decreasing when $z_5>0$, so its maximum value
is attained in the region $z_5\leq 0$. To obtain the largest possible value of $w_5$, we need to ensure that $\dot{w}_5$
is as large as possible in this region. In view of \eqref{4-m}, this simply means that equality must hold in the first
inequality in \eqref{5-as}.  We are thus faced with a situation which is almost identical with \eqref{3-as}.  Arguing as
before, we find that the curve must satisfy
\begin{equation*}
2\dot{x} = y-x, \quad\quad (5x+y-4) \dot{x} + 2(1-x) \dot{y} = -\frac{z_5(x,y)}{2}
\end{equation*}
until it exits the region $z_5\leq 0$.  This is the same system of ODEs that we had in \eqref{ode}, and the rest of our
argument applies almost verbatim.  The solution associated with the unstable manifold at the origin intersects the curve
$z_5=0$ at the point $(0.2746,0.6180)$ and so 
\begin{equation*}
w_5(0.2746, 0.6180) \approx 37.924
\end{equation*}
is the largest possible value of $w_5$. Using this fact, we get the upper bound on $x$ which is stated in the theorem.
To show that our results for this case are sharp, we follow our approach in the isotropic case with dominant energy.  As
there are only minor changes that need to be made, we are going to omit the details.

\section*{Acknowledgements}
We would like to thank Aur\'{e}lien Decelle to whom we are indebted for both the numerical analysis and the figures which
appear in this paper.  We would also like to thank Petros Florides for his encouragement and H\aa{}kan Andr\'{e}asson
whose recent papers \cite{Hak2, Hak3, Hak} have revived interest in this important problem.

\end{document}